\documentclass[12pt]{iopart}
\usepackage{graphicx}
\usepackage{bm}
\usepackage{times}
\usepackage[hypertex]{hyperref}
\usepackage{slashed}
\newcommand{\be}{\begin{equation}}
\newcommand{\ee}{\end{equation}}
\newcommand{\beq}{\begin{eqnarray}}
\newcommand{\eeq}{\end{eqnarray}}
\newcommand{\ba}{\begin{array}}
\newcommand{\ea}{\end{array}}

\begin{document}

\title[Scalar Bilepton Dark Matter]{Scalar Bilepton Dark Matter}

\author{C. A. de S. Pires and P. S. Rodrigues da Silva}
\address{Departamento de F\'{\i}sica, Universidade Federal da
Para\'{\i}ba, \\ 
Caixa Postal 5008, 58051-970, Jo\~ao Pessoa - PB, Brazil.}

\ead{cpires@fisica.ufpb.br and psilva@fisica.ufpb.br}

\date{\today}

\begin{abstract}
In this work we show that 3-3-1 model with right-handed neutrinos has a natural weakly interacting massive particle (WIMP) dark mater candidate. It is a complex scalar with mass of order of some hundreds of GeV which carries two units of lepton number, a scalar bilepton. This makes it a very peculiar WIMP, very distinct from Supersymmetric or Extra-dimension candidates. Besides, although we have to make some reasonable assumptions concerning the several parameters in the model, no fine tunning is required in order to get the correct dark matter abundance. We also analyze the prospects for WIMP direct detection by considering recent and projected sensitivities for WIMP-nucleon elastic cross section from CDMS and XENON Collaborations. 
\end{abstract} 


\section{Introduction} \setcounter{equation}{0}
\label{sec1}
The problem of matter density in the Universe seems to be one of the most intriguing and exciting subjects in modern Physics. The growing refinement achieved in cosmological data leaves no doubt about a dark component in the observed mass density, constituting roughly $22\%$ of all energy density acording to the three year run of WMAP~\cite{WMAP3}. This yet unknown component has to be non-baryonic, its interaction with the electroweak Standard Model (SM) particles should be negligible  and it has to be cold, i.e., non-relativistic at the time it decouples from the radiation bath, the so called  Cold Dark Matter (CDM).
From the theoretical side, there are some proposals to explain the CDM in the context of Particle Physics models (see Ref.~\cite{DMmodels,Murayama,dmoutros,Dobrescu,little} and references therein for a review of the subject). Among them there are models which present natural candidates to play this role, the weakly interacting massive particles (WIMP)'s, with mass ranging from approximately $1$~GeV to $1$~TeV. These WIMP's are nice candidates because their masses are in the GeV realm, turning them cold at decoupling, and mainly because their weakly interacting aspect not only yields a thermally averaged annihilation cross section of order of weak interactions, leading to the expected order of magnitude to CDM abundance, but also coincides with the scale of Particle Physics models to be probed at the Large Hadron Collider (LHC) that finds itself at the final stage to start its running phase~\cite{LHC}. It also presents the possibility of being seen in direct detection experiments since its massiveness would imply an observable recoil of nuclei in elastic collisions~\cite{DMmodels,wimpexperiments}.

The most promising scenarios where such WIMP's can be present in the particle spectrum are Supersymmetry (SUSY) and Extra Dimensions models~\cite{DMmodels,Murayama,Dobrescu}. All these models dispose of some kind of discrete symmetry in order to stabilize their CDM candidates. Also, they have to be realized at the electroweak scale so that their new particles are potentially good candidates for CDM. 
Although such models may represent the greatest expectations in Particle Physics for the Physics at TeV scale to be probed by LHC, the absence of any experimental evidence to support these models allows us to work with alternatives.
One of such alternative routes concerns the enlargement of the gauge symmetry group from $SU_C(3)\otimes SU_L(2)\otimes U_Y(1)$ to larger groups. In particular, there exists a simple extension of the SM gauge group to $SU_C(3)\otimes SU_L(3)\otimes U_X(1)$, the so called 3-3-1 model~\cite{PleitezPisano}.

This class of models is interesting for several reasons, not only because they mean a different scenario, but because they possess several nice features. For example, (i) the family problem is absent in this model since it demands that there be only three families of fermions when anomalies are canceled and asymptotic freedom is considered~\cite{PleitezPisano}; (ii) electric charge quantization is automatic~\cite{qcharge}; (iii) right-handed neutrinos can be part of the spectrum in some versions of the model~\cite{331valle,331RH,singer} and their tiny observed mass difference can be easily accommodated~\cite{lightnu}; (iv) axions and majorons are a natural outcome in some versions~\cite{axionmajoron}, providing light particles which could also contribute to the problem of dark matter origin. Besides, it is possible that a custodial symmetry exists in these models which would make them indistinguishable from SM at low energy scales~\cite{331np}, and it would then be a strong rival to SM itself.

In this work we will concentrate on the 3-3-1 version of the model with right handed neutrinos in the spectrum (3-3-1$RH\nu$) model~\cite{331RH}. The reason behind this choice relies on the fact that neutrinos mass is already a mandatory property that needs to be included in all reasonable extensions of SM. Besides, the model can be implemented with just three scalar triplets instead of including a sextet as in other versions, being economical in its content. However, the most appealing motivation to deal with 3-3-1$RH\nu$ to explain the origin of CDM is due to the possibility of having a candidate with a very distinct signature.
Among its properties the model can be made lepton number conserving if some of its fields carry two units of lepton number which will be called bileptons. This peculiar property has many phenomenological implications. Namely, rare lepton decays can emerge, neutrinoless double beta decay is allowed, right handed neutrinos are going to appear as byproducts of heavy vector bileptons decays and so on. 
It is then automatic to ask if some of these additional bilepton fields can be a CDM candidate, once it is provided with a very specific quantum number appropriate to forbid its interaction with many of the electroweak fields. This would play a similar role as that played by the discrete symmetries in the competing models cited above. Moreover, as the sought candidate is merged in the exotic new effects just mentioned, their appearance in the coming collider experiments would represent an unquestionable evidence of our CDM candidate.

What we are going to investigate in this work is the possible realization of this scenario in the 3-3-1$RH\nu$ model for one of the bilepton scalars. We are interested in identifying this field and characterize it as a WIMP. This can be realized if it can be shown that it is stable in the range of parameters for which the abundance is in accordance with the recent data of WMAP~\cite{WMAP3}. It is important to stress that the natural perturbative scale for the 3-3-1 models is on the TeV scale~\cite{alexlimit}, which is suitable for obtaining a WIMP. In view of all this, it seems that such a possibility is as welcome as any previous attempt to explain the CDM content through SUSY or  Extra dimensions, offering a completely new and distinct particle to do this job. It should be mentioned that other works exist in the literature trying to explain CDM in the context of 3-3-1 models~\cite{DM331}, but their aim was to obtain a self-interacting dark matter to avoid excessively dense cores in the center of galaxies and clusters as well as excessive large number of halos within the local group when contrasted to observations~\cite{spergel}, which demands a light dark matter candidate. We do not pursue this approach here. On the contrary, we want to show that the 3-3-1$RH\nu$ model possesses a bilepton scalar that can play the role of a WIMP, which is the preferred candidate for CDM. Besides, we get this without the need to fine tune its couplings to small and unnatural values.

This work is organized as follows. In Sec.~\ref{sec2} we present the model with its content and interactions. In Sec.~\ref{sec3} we diagonalize the mass matrices for the particle spectrum, allowing us to characterize the new extra particles. In Sec.~\ref{sec4} we identify our WIMP with one of the neutral scalar bileptons of the model, checking its viability as CDM candidate by computing its abundance for a range of values of the parameters, which turns out to be natural in the sense we do not need to make any fine adjustment on the parameters. Then we study the prospects of direct detection for this WIMP. We conclude with Sec.~\ref{sec5}.

\section{The model}
\label{sec2}
In the 3-3-1RH$\nu$ model the leptons come in triplet and in singlet representations,
\begin{eqnarray}
f_{aL} = \left (
\begin{array}{c}
\nu_{aL} \\
e_{aL} \\
\nu^C_{aL}
\end{array}
\right )\sim(1\,,\,3\,,\,-1/3)\,,\,\,\,e_{aR}\,\sim(1,1,-1),
 \end{eqnarray}
while in the quark sector, one generation comes in the triplet and the other two compose
an anti-triplet representation with the following content,
\begin{eqnarray}
&&Q_{iL} = \left (
\begin{array}{c}
d_{i} \\
-u_{i} \\
d^{\prime}_{i}
\end{array}
\right )_L\sim(3\,,\,\bar{3}\,,\,0)\,,u_{iR}\,\sim(3,1,2/3),\,\,\,\nonumber \\
&&\,\,d_{iR}\,\sim(3,1,-1/3)\,,\,\,\,\, d^{\prime}_{iR}\,\sim(3,1,-1/3),\nonumber \\
&&Q_{3L} = \left (
\begin{array}{c}
u_{3} \\
d_{3} \\
u^{\prime}_{3}
\end{array}
\right )_L\sim(3\,,\,3\,,\,1/3),u_{3R}\,\sim(3,1,2/3),\nonumber \\
&&\,\,d_{3R}\,\sim(3,1,-1/3)\,,\,u^{\prime}_{3R}\,\sim(3,1,2/3)
\label{quarks} 
\end{eqnarray}
where $a = 1,\,2,\, 3$ refers to the three generations and the index $i=1,2$ is restricted to only two generations. The primed quarks
are new heavy quarks with the usual electric charges. Actually,  as we will see below, they are leptoquarks since besides baryon number they also carry two units of lepton number.

In order to generate the right masses for the known particles
the model requires only  three scalar triplets, namely,
\begin{eqnarray}
 \chi = \left (
\begin{array}{c}
\chi^0 \\
\chi^{-} \\
\chi^{\prime 0}
\end{array}
\right ),\,
\eta = \left (
\begin{array}{c}
\eta^0 \\
\eta^- \\
\eta^{\prime 0}
\end{array}
\right ),\,\rho = \left (
\begin{array}{c}
\rho^+ \\
\rho^0 \\
\rho^{\prime +}
\end{array}
\right ) 
, \label{scalarcont} 
\end{eqnarray}
with $\eta$ and $\chi$ both transforming as $(1\,,\,3\,,\,-1/3)$
and $\rho$ transforming as $(1\,,\,3\,,\,2/3)$.

We assume the following discrete symmetry transformation for the fields in order to have a minimal model,
\be
\left( \chi\,,\,\rho\,,e_{aR}\,,\, u_{aR}\,,\,u^{\prime}_{3R}\,,\,d^{\prime}_{iR}\,,\, Q_{3L} \right) \rightarrow -\left( \chi\,,\,\rho\,,e_{aR}\,,\, u_{aR}\,,\,u^{\prime}_{3R}\,,\,d^{\prime}_{iR}\,,\, Q_{3L}\right)
	\label{discretesymmetryI}
\ee
This symmetry helps in avoiding the undesirable Dirac mass terms for the neutrinos and  allows a minimal charge conjugation and parity (CP) conserving potential~\cite{SCPV},
\begin{eqnarray} V(\eta,\rho,\chi)&=&\mu_\chi^2 \chi^2 +\mu_\eta^2\eta^2
+\mu_\rho^2\rho^2+\lambda_1\chi^4 +\lambda_2\eta^4
+\lambda_3\rho^4+ \nonumber \\
&&\lambda_4(\chi^{\dagger}\chi)(\eta^{\dagger}\eta)
+\lambda_5(\chi^{\dagger}\chi)(\rho^{\dagger}\rho)+\lambda_6
(\eta^{\dagger}\eta)(\rho^{\dagger}\rho)+ \nonumber \\
&&\lambda_7(\chi^{\dagger}\eta)(\eta^{\dagger}\chi)
+\lambda_8(\chi^{\dagger}\rho)(\rho^{\dagger}\chi)+\lambda_9
(\eta^{\dagger}\rho)(\rho^{\dagger}\eta) \nonumber \\
&&-\frac{f}{\sqrt{2}}\epsilon^{ijk}\eta_i \rho_j \chi_k +\mbox{H.c}.
\label{potential}
\end{eqnarray}
Also, the Yukawa sector can be written as,
\begin{eqnarray}
&-&{\cal L}^Y =f_{ij} \bar Q_{iL}\chi^* d^{\prime}_{jR} +f_{33} \bar Q_{3L}\chi u^{\prime}_{3R} + g_{ia}\bar Q_{iL}\eta^* d_{aR} \nonumber \\
&&+h_{3a} \bar Q_{3L}\eta u_{aR} +g_{3a}\bar Q_{3L}\rho d_{aR}+h_{ia}\bar Q_{iL}\rho^* u_{aR}+ G_{aa}\bar f_{aL} \rho e_{aR}+\mbox{H.c}. 
\label{yukawa}
\end{eqnarray}
With this set of Yukawa interactions all fermions, except the neutrinos, gain mass. In this model neutrinos masses are generated through effective dimension-five operators as shown in Ref.~\cite{lightnu}.

In the gauge sector, the model recovers the usual SM gauge bosons, $W^{\pm}\,,\,Z^0\,,\, \gamma$,  and contains five additional vector bosons called  $V^{\pm}$, $U^0$, $U^{0 \dagger}$ and $Z^{\prime}$~\cite{331RH}, with masses around hundreds of GeV.

Considering the required properties for a dark matter candidate, it will be important to have in mind that some  of the  new particles  carry two units of lepton number {\bf L} and are known as bileptons, namely,
\begin{eqnarray} {\mbox {\bf L}}(V^+\,,\, U^{\dagger0}\,,\, u^{\prime}_{3} \,,\, \eta^{\prime
0}\,,\,\rho^{\prime +})=-2 \,,\,\,\,\,\,{\mbox {\bf L}}(V^- \,,\,U^0 \,,\,d^\prime_{i}
\,,\, \chi^ 0\,,\, \chi^-)=+2. \label{leptonnumber} \end{eqnarray}
This assignment is such that the lagrangian is lepton number conserving. Observe that this quantum nunber association limits the range of interactions available to the bileptons since the only gauge bosons that carry lepton number are the new ones, and their masses are at the TeV scale. This property will show itself useful when considering the WIMP stability in Sec.~\ref{sec4}. 

\section{The mass eigenstates}
\label{sec3}
From the scalar content of the model Eq.~(\ref{scalarcont}) we have five neutral scalars at our disposal. Two of them carry two units of lepton number, $\eta^{\prime 0}\,,\,\chi^0$ and if lepton number is conserved (as we assume here) they do not develop vacuum expectation value (VEV).  As for the remaining ones, since they do not carry lepton number they are free to develop nontrivial VEV's,
\begin{eqnarray}
 \eta^0 , \rho^0 , \chi^{\prime 0} \rightarrow  \frac{1}{\sqrt{2}} (v_{\eta ,\rho ,\chi^{\prime}} 
+R_{ \eta ,\rho ,\chi^{\prime}} +iI_{\eta ,\rho ,\chi^{\prime}}). 
\label{vacua} 
\end{eqnarray} 

On substituting this expansion in the above potential, we obtain the following set of constraints, 
\begin{eqnarray} &&\mu^2_\chi +\lambda_1 v^2_{\chi^{\prime}} +
\frac{\lambda_4}{2}v^2_\eta  +
\frac{\lambda_5}{2}v^2_\rho-\frac{f}{2}\frac{v_\eta v_\rho}
{ v_{\chi^{\prime}}}=0,\nonumber \\
&&\mu^2_\eta +\lambda_2 v^2_\eta +
\frac{\lambda_4}{2} v^2_{\chi^{\prime}}
 +\frac{\lambda_6}{2}v^2_\rho -\frac{f}{2}\frac{v_{\chi^{\prime}} v_\rho}
{ v_\eta} =0,
\nonumber \\
&&
\mu^2_\rho +\lambda_3 v^2_\rho + \frac{\lambda_5}{2}
v^2_{\chi^{\prime}} +\frac{\lambda_6}{2}
v^2_\eta-\frac{f}{2}\frac{v_\eta v_{\chi^{\prime}}}{v_\rho} =0.
\label{mincond} \end{eqnarray}
It is reasonable to assume that $f$ is of the order of $v_{\chi^{\prime}}$, the scale associated to 3-3-1 breaking ocurring at few TeV, while $v_\eta$ and $v_\rho$ are related to electroweak symmetry breaking. In what follows we assume, for simplicity, that the VEV's related to the ordinary vector bosons mass are the same, and use the convenient notation, $v_\eta =v_\rho\equiv v $,  and  also $f=v_{\chi^{\prime}}/2\equiv V/2$.

By construction, lepton number is conserved by the interactions of this model, implying that the neutral scalars $( \chi^0\,,\,\eta^{\prime 0})$  do not mix with the other three $\chi^{\prime 0}\,,\,\eta^0\,,\,\rho^0$.  Thus, in the basis $( \chi^0\,,\,\eta^{\prime 0})$ we have the following mass matrix~\footnote{Here we are considering the CP even and CP odd scalars altogether, $\chi^0 = R_{\chi^0}+ i I_{\chi^0}$ and  $\eta^{\prime 0} = R_{\eta^{\prime 0}}+ i I_{\eta^{\prime 0}}$, since they possess the same mass matrices. },
\begin{eqnarray}
 \frac{V^2}{4}(\lambda_7 +1/2)\left(\begin{array}{cc} 
   \frac{v^2}{V^2}  & \frac{v}{V}  \\
  \frac{v}{V}  & 1
  \end{array}
\right). \label{massmatrixchieta} 
\end{eqnarray}
After we diagonalize this matrix, we obtain a zero mass scalar which is given by
\beq
G=-\frac{V}{v\sqrt{\frac{V^2}{v^2}+1}}\chi^0+\frac{1}{\sqrt{\frac{V^2}{v^2}+1}}\eta^{\prime 0}\,,
\label{G1}
\eeq
recognized as the Goldstone boson eaten by the gauge bosons $U^0$  and $U^{0 \dagger}$. The other scalar, 
\beq
\phi= \frac{1}{\sqrt{\frac{V^2}{v^2}+1}}\chi^0 +\frac{V}{v\sqrt{\frac{V^2}{v^2}+1}}\eta^{\prime 0}\,,
\label{wimp}
\eeq
is a heavy scalar with mass given by $M^2_{\phi}=\frac{1}{4}(\lambda_7+\frac{1}{2})(v^2+V^2)$. The assumption $V>>v$, allows us to say that the Goldstone bosons are mostly contained in the complex scalar $G \approx \chi^0$, while the heavy scalar is mostly  $\phi \approx \eta^{\prime 0}$. We are going to show in the next section that $\phi$ has all the appropriate features to be our WIMP candidate for dark matter. Meanwhile we perform the mass matrix diagonalization for the remaining neutral scalars as well as the charged ones. We do that because we need to have control of all masses in the model in order to guarantee the $\phi$ stability.   

Considering the expansion in Eq.~(\ref{vacua}), we have the following mass matrix for the CP-even scalars in the basis $(R_{\chi^{\prime}}\,,\,R_\eta\,,\,R_\rho)$, 
\begin{eqnarray}
 V^2\left(\begin{array}{ccc} 
    \lambda_1+\frac{1}{8}\frac{v^2}{V^2} & (\frac{\lambda_4}{2}-\frac{1}{8})\frac{v}{V}& (\frac{\lambda_5}{2}-\frac{1}{8})\frac{v}{V} \\
  (\frac{\lambda_4}{2}-\frac{1}{8})\frac{v}{V}  & \frac{1}{8}+\lambda_2\frac{v^2}{V^2} & -\frac{1}{8}+\frac{\lambda_6}{2}\frac{v^2}{V^2} \\
  (\frac{\lambda_5}{2}-\frac{1}{8})\frac{v}{V} & -\frac{1}{8}+\frac{\lambda_6}{2}\frac{v^2}{V^2} & \frac{1}{8}+\lambda_3 \frac{v^2}{V^2}
  \end{array}
\right). \label{massmatrixR} 
\end{eqnarray}
Its diagonalization leads to the following scalar eigenvectors, 
\beq
h^0_1   &\approx & R_{\chi^{\prime}}\,,\nonumber \\ \nonumber
 h^0_2 &\approx &  \frac{1}{\sqrt{2}}(R_\eta-R_\rho)\,,\\ \nonumber
 H  &\approx &  \frac{1}{\sqrt{2}}(R_\eta+R_\rho)\,,
\label{eigenvextorsRe}
\eeq
whose masses are respectively,
\beq
m^2_{h^0_1}  &\approx &  \lambda_1 V^2 +\frac{v^2}{8}\,,\nonumber \\ \nonumber
m^2_{h^0_2} &\approx &   \frac{1}{8}\left[(4\lambda_2+4\lambda_3)v^2+V^2\left(1+\sqrt{1-8\lambda_6\frac{v^2}{V^2}}\right)\right]\,,\nonumber \\
m^2_{H} &\approx &   \frac{1}{8}\left[(4\lambda_2+4\lambda_3)v^2+V^2\left(1-\sqrt{1-8\lambda_6\frac{v^2}{V^2}}\right)\right]\,.
\label{eingenvaluesRe}
\eeq
Among these, the first and second scalars in the above equation are heavy while the third one  is the lightest scalar of the theory, which we recognize as the standard Higgs boson.

Regarding the CP-odd scalars, we have the following mass matrix in the basis $(I_{\chi^{\prime}}\,,\,I_\eta\,,\,I_\rho)$,
\begin{eqnarray}
 \frac{V^2}{8}\left(\begin{array}{ccc} 
  \frac{v^2}{V^2}   &\frac{ v}{V} & \frac{v}{V} \\
  \frac{ v}{V}  & 1 & 1 \\
  \frac{ v}{V} & 1 & 1
  \end{array}
\right). \label{massmatrixI} 
\end{eqnarray}
This matrix has the following eigenvectors,
\beq
I^0_1 &\approx & I_{\chi^{\prime}}\,, \nonumber \\
I^0_2 &\approx & \frac{1}{\sqrt{2}}(I_\rho-I_\eta)\,,  \nonumber \\
I^0_3 &\approx & \frac{1}{\sqrt{2}}(I_\rho+I_\eta)\, \nonumber \\
\label{vetoresCPodd}
\eeq
where $I^0_1$ and $I^0_2$ correspond to null eigenvalues and $I^0_3$ is the massive pseudo-scalar with
\beq
m_{I^0_3}&=& \frac{1}{4}(V^2+\frac{v^2}{2})\,.
\label{autovalorCPodd}
\eeq
The pseudo-scalars, $I^0_1$  and $I^0_2$, are identified as the Goldstones eaten by the neutral gauge bosons, the standard $Z^0$ and the   $Z^{\prime}$ characteristic of the extended 3-3-1 gauge symmetry.

Finally, let us consider the charged scalar mass matrices. Remember that two of the charged scalars carry two units of lepton number (charged scalar bileptons), namely, $\chi^-\,,\,\rho^{\prime -}$, while the other two, $\eta^- \,,\, \rho^-$, have zero lepton number. This means that the first two charged scalars do not mix with the last two. Thus we are going to have two $2\times2$ mass matrices. The first one, in the basis $(\chi^-\,,\,\rho^{\prime -})$ takes the form,
\begin{eqnarray}
 \frac{V^2}{2}(\lambda_8 +1/2)\left(\begin{array}{cc} 
   \frac{v^2}{V^2}  & \frac{v}{V}  \\
  \frac{v}{V}  & 1
  \end{array}
\right). \label{massmatrixcharged1} 
\end{eqnarray}
The other mass matrix of charged scalars, in the basis $(\eta^-\,,\,\rho^{ -})$, is given by
\begin{eqnarray}
 (\frac{\lambda_9}{2} v^2 +\frac{1}{4}V^2)\left(\begin{array}{cc} 
   1  & 1  \\
  1  & 1
  \end{array}
\right). \label{massmatrixcharged2} 
\end{eqnarray}
Their diagonalization shows that two of the four eigenvalues are null and correspond to the eigenvectors,
\beq
h^-_1 &\approx & \chi^-\,, \nonumber \\
h^-_2 &\approx & \frac{1}{\sqrt{2}}(\eta^- -\rho^-)\,,
\label{vetoresGC}
\eeq
while the massive states are,
\beq
h^-_3 &\approx & \rho^{- \prime }\,, \nonumber \\
h^-_4 &\approx & \frac{1}{\sqrt{2}}(\eta^-+\rho^-)\,.
\label{vetoresCS}
\eeq
Again, the massless charged scalars are Goldstones, $h^-_1$ is eaten by the charged gauge boson $V^{-}$ (recall that both are bileptons), and $h^-_2$ is the Goldstone eaten by the
standard gauge boson $W^{-}$. The charged scalars remaining in the spectrum, $h^-_3$ (bilepton) and $h^-_4$, possess the following masses,
\beq
m^2_{h^-_3} &=& \frac{1}{2}(\lambda_8+\frac{1}{2})(V^2+v^2)\,,
\nonumber \\
m^2_{h^-_4} &=& \frac{V^2}{2}+\lambda_9 v^2\,.
\label{massCS}
\eeq
This completes our scalar bosons analysis, assuring us of the correct number of Goldstones, the identification of the Higgs boson and the extra heavy scalars. We observe that we can choose the scalar coupling constants of order 0.1\footnote{Some of these couplings can be adjusted around this value in order to yield a Higgs mass of 114~GeV or a little higher, but no fine tunning is required for that.} to avoid any fine tunning. Also, this choice turns the potential stable and all eigenvalues of the mass matrices positive. Finally, to guarantee that $\phi$ be the lightest new particle in the spectrum we have only to assume that $\lambda_7$ is negative, which does not alter the stability of the potential and proves to be a choice as good as a positive coupling.

Bellow we present the vector bosons masses of 3-3-1$RH_\nu$ model, so that we can compare them with the scalar bosons. We assume the experimental values for the SM gauge boson $Z$, $M_Z=91.118$~GeV, while 
\beq
m_{W^\pm}^2&=& \frac{1}{2}g^2v^2\,,
\nonumber \\
m^2_{V^\pm} &=& m^2_{U^0} = \frac{1}{4}g^2(V^2+v^2)\,,
\nonumber \\
m^2_{Z^\prime} &=& \frac{1}{3-4s_W^2}g^2(V^2+v^2)\,,
\label{massvec}
\eeq
where we have used the approximation $V\gg v$ and neglected terms of order higher than ${\cal O}(v^2/V^2)$. The above degeneracy for the vector bileptons is due to our simplifying choice $v_\eta = v_\rho \equiv v$.

Next we expose our reasons to choose $\phi=\eta^{\prime 0}$ as our candidate for the 3-3-1 WIMP. 

\section{The 3-3-1 WIMP}
\label{sec4}

A good CDM candidate should be an electrically neutral particle, be stable and generate an energy density about 22$\%$ of the critical energy density in the universe.  The 3-3-1RH$\nu$ model particle which fulfils these criteria, as we next show, is the complex scalar $\phi$, which is electrically neutral, can be stable if it is the lightest new particle in the spectrum and, for some range of the parameters, can be shown to possess the right abundance for CDM. 

Concerning $\phi$ stability, notice that it carries two units of lepton number {\bf L} and since the model is lepton number conserving, its decay can proceed only through final states with the same total {\bf L}. 
Then $\phi$ triple interactions can be generally cast as $\Gamma_{\mbox{triple}}=(\phi)\times(A)\times (B$), where $A$ is any particle carrying two units of lepton number, as given in Eq.~(\ref{leptonnumber}), and $B$ is any particle possessing null lepton number. By assuring that particles carrying {\bf L}$=\pm2$ which interact with $\phi$ are heavier than $\phi$, its stability is then secured. 

By considering the Yukawa interactions Eq.~(\ref{yukawa}), we see that $\phi$ does not couple to leptons, but does couple to the new heavy quarks, which also carry two units of lepton number, $u^{\prime}$,  $d^{\prime}$ or $s^\prime$, plus an ordinary quark. The bilepton quarks are expected to be heavier than $\phi$ since their masses are proportional to V, being of order of few TeV, thus forbidding $\phi$ to decay into fermions. As for the vector bosons as final states, once the decay has to involve $V^{\pm}$ or $U^0$ to conserve lepton number, it is forbidden if $\phi$ is lighter than $V^{\pm}$ or $U^0$. This will be the case for almost the whole range of values of the free parameters which lead to the correct abundance and Higgs mass, as long as $\lambda_7 < 0$ in the potential Eq.~(\ref{potential}).
Also, from the potential Eq.~(\ref{potential}) and the lepton number assignment Eq.~(\ref{leptonnumber}), all triple scalar interactions with $\phi$ involve simultaneously $h^\pm_3$ and $h^\pm_4$, in order to guarantee lepton number and charge conservation. However, as can be seen from Sec.~\ref{sec3}, all scalars except the Higgs are heavier than $\phi$ given that $\lambda_7 < 0$ and the other couplings are of order of one~\footnote{See also in Fig.~\ref{fig:5} the mass of the particles which interact with $\phi$ as a function of the free parameters relevant for this work.}, thus $\phi$ cannot decay into a pair of scalars.
These are all interactions we need to consider to be sure that $\phi$ is stable. 

Next, when computing the thermal averaged cross section, we establish the set of parameters that realizes this scenario. 

\subsection{Relic abundance}
\label{subrelic}

Once we have identified which particle can be our WIMP CDM candidate, it is imperative to obtain the correct observed CDM abundance $\Omega_{CDM}$. 
We consider that the WIMP is in thermal equilibrium with radiation in the early epochs until its rate of reaction becomes smaller than the rate of expansion of the Universe, i.e. it decouples from thermal bath, the so called freeze-out. In the case of WIMPs this happens when they are non-relativistic.   
The Boltzmann equation dictates the evolution of the particle number density $n$ with the expanding Universe,
\be
\frac{dn}{dt}+3Hn = -<\sigma v_{rel}>(n^2-n^2_{eq}),
\label{boltzmann}
\ee
where, $H$ is the Hubble parameter or expansion rate of the Universe, which can be written as $H^2=8\pi\rho/3M_{Pl}^2$ for a flat Universe, $<\sigma v_{rel}>$ is the thermal averaged cross section for WIMP annihilation times the relative velocity, $n_{eq}$ is the particle number density at equilibrium and $M_{Pl}\approx 1\times 10^{19}$~GeV is the Planck mass and $\rho$ is the energy density of the Universe. Since WIMP is non-relativistic at the time of decoupling ($T<<M_{\phi}$), its equilibrium number density is,
\be
n_{eq}=g_\phi\left(\frac{M_{\phi}T}{2\pi}\right)^\frac{3}{2}e^{-\frac{M_{\phi}}{T}},
\label{nequilibrium}
\ee
with $g_\phi$ the WIMP number of degrees of freedom. Next we follow the standard procedure derived in Ref.~\cite{DMmodels,SWO} to obtain a solution to the above equation and then determine the WIMP abundance. By defining $x\equiv M_{\phi}/T$ and using the non-relativistic approximation for the squared center of mass energy, $s=4M_\phi^2+M_\phi^2v^2$, we expand the cross section till the first power in $v^2$, resulting in the thermally averaged cross section,
\be
<\sigma v_{rel}>\approx a+\frac{6b}{x},
\label{thermalexpansion}
\ee
where $a$ and $b$ are the model dependent parameters. The relic WIMP abundance in the context of a flat cosmological constant dominated Universe ($\Lambda$CDM model), can be expressed as,
\be
\Omega_\phi h^2 \approx \frac{1.04\times10^9}{M_{Pl}}\frac{x_F}{\sqrt{g^*}(a+\frac{3b}{x_F})}\,,
\label{omega}
\ee
where $x_F$ is $x\equiv M_{\phi}/T$ computed at the temperature of decoupling of dark matter from equilibrium, the freeze-out temperature given by,
\be
x_F = \ln{\left[c(c+2)\sqrt{\frac{45}{8}}\frac{g_\phi}{2\pi^3}\, \frac{M_\phi\, M_{Pl}\, (a+\frac{6b}{x_F})}{\sqrt{g^* x_F}}\right]}\,,
\label{xf}
\ee
which in general is close to $x_F\approx 20$.
Also, $g^*$ is the number of degrees of freedom of relativistic particles in thermal equilibrium with the WIMP at freeze-out, which in our case is about $g^*=183/2=91.5$ and  $g_\phi=2$. The constant $c$ of order of one is obtained by matching the late-time and early-time solutions for the abundance and, for our purposes, it is enough to take it as $c=1/2$ since it has only a small effect in the logarithmic dependence of $x_F$.

We can solve Eq.~\ref{xf} iteratively and plug the result in Eq.~\ref{omega} in order to compare the WIMP abundance with the latest results of WMAP~\cite{WMAP3} which, according to the $\Lambda$CDM model, imposes the following bounds to the dark matter abundance at $2\sigma$ level,
\be
0.096 < \Omega_{DM}h^2 < 0.122\,.
\label{limitsDM}
\ee

\begin{figure}[h]
\centering
\includegraphics[width=0.7\columnwidth]{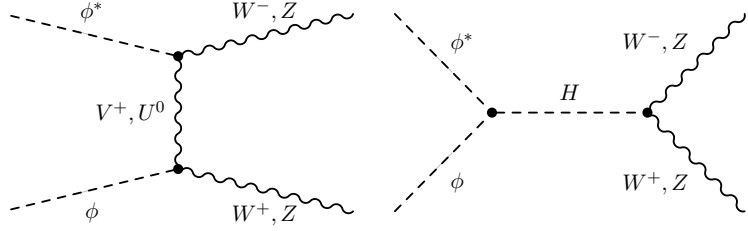}
\caption{WIMP annihilation into Standard Model gauge bosons, $\phi^*\phi\rightarrow W^+W^-$ and $\phi^*\phi\rightarrow ZZ$.}
\label{fig:1}
\end{figure}

\begin{figure}[h]
\centering
\includegraphics[width=0.9\columnwidth]{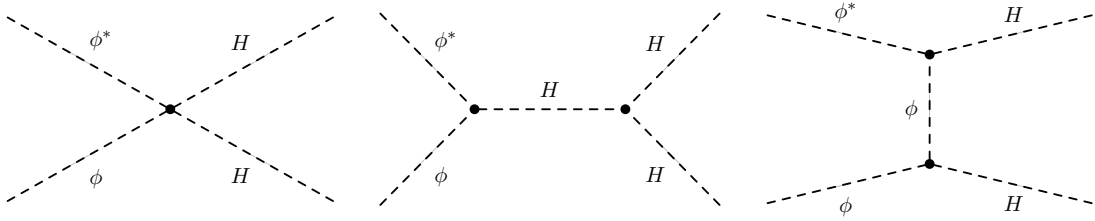}
\caption{WIMP annihilation into Standard Model Higgs, $\phi^*\phi\rightarrow HH$.}
\label{fig:2}
\end{figure}

\begin{figure}[h]
\centering
\includegraphics[width=0.7\columnwidth]{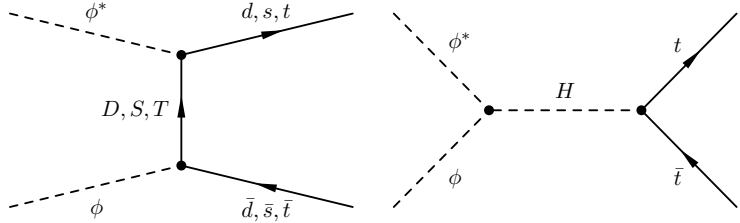}
\caption{WIMP annihilation into Standard Model quarks, $\phi^*\phi\rightarrow \bar{q}q$.}
\label{fig:3}
\end{figure}

The processes contributing to $<\sigma v_{rel}>$ are mainly the annihilation into gauge bosons, $\phi\phi\rightarrow W^+ W^-$ and $\phi\phi\rightarrow Z Z$ depicted in Fig.~\ref{fig:1}, the annihilation into Higgs boson, $\phi\phi\rightarrow H H$ shown in Fig.~\ref{fig:2}, which is comparable to the gauge bosons contributions, and the annihilations into quarks $\phi\phi\rightarrow \bar{q} q$ presented in Fig.~\ref{fig:3}, whose role is better discussed bellow.

Our first step to calculate the thermally averaged cross section parameters, $a$ and $b$ appearing in the above equations, is to set the range of free parameters of the model. In order to proceed with this we make some simplifying assumptions since there are too many free parameters to deal with.
A natural choice is to take the couplings not far from ${\cal O}(1)$. In this sense, and considering the relevant interactions (see the appendix) involved in these annihilation channels, we leave the couplings, $\lambda_1$ and $\lambda_2$, free and fix 
\beq
\lambda_3&=&\lambda_4=\lambda_5=\lambda_8=\lambda_9=\lambda_1\,, \nonumber \\
\lambda_6&=&-\lambda_7=\lambda_2\,.
\label{couplings}
\eeq 
This convenient set of parameters allows us to vary the Higgs and $\phi$ masses as we wish and the only care we have to take is not to make the bilepton gauge bosons, $V^\pm$ and $U^0$, lighter than $\phi$. As we will see next, this is the case if $V > 1.2$~TeV and $\lambda_2 > 0.078$, which is true for all parameter space which is compatible with WMAP results for $\Omega_{DM}$.

Concerning the annihilation into quarks, we will assume that the Yukawa couplings are dominant for the quarks in the same family and much smaller otherwise. These assumptions allow us to neglect any mixing among quarks in this study. In this case, according to Eq.(\ref{yukawa}), the relevant couplings are given in terms of down, strange and top mass, $g_{11}=\sqrt{2}m_d/v_\eta$, $g_{22}=\sqrt{2}m_s/v_\eta$ and $h_{33}=\sqrt{2}m_t/v_\eta$, respectively. The WIMP annihilation into quarks are restrict then to the following reactions,
\be
\phi\phi \rightarrow \bar{d}d\,,\,\,\,\,\,\,\,
\phi\phi \rightarrow \bar{s}s\,,\,\,\,\,\,\,\,
\phi\phi \rightarrow \bar{t}t\,.
\label{quarkscontrib}
\ee
It should be stressed that although the annihilation into top quark is the only one where the s channel can be dominant since it proceeds by  an intermediate Higgs, the same is not true for annihilation into down or strange quarks.
This happens because the exchange channels are not suppressed by the lightness of quark mass as happens for the s channel. Indeed, for the Yukawa couplings assumed here, the amplitudes for t and u channels are some orders of magnitude higher than the amplitude for the s channel in the process $\phi\phi \rightarrow \bar{d}d$ and $\phi\phi \rightarrow\bar{s}s$. Nevertheless, when we compare the magnitude of these contributions with the s channel amplitude for  $\phi\phi \rightarrow \bar{t}t$, they are negligible. For this reason we only take into account those diagrams in Fig.~\ref{fig:3} for top quark annihilation.

\begin{figure}[h]
\centering
\includegraphics[width=0.7\columnwidth]{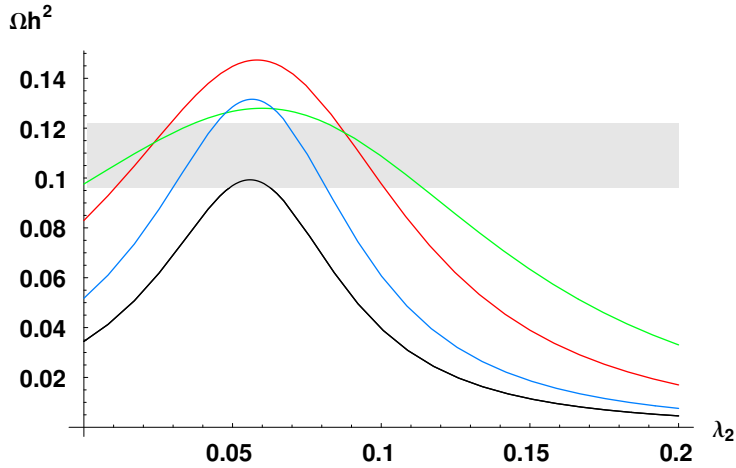}
\caption{The WIMP abundance as a function of the coupling $\lambda_2$, with $\lambda_1=0.7$ and other couplings constrained to these ones as exposed in the text. 
From the right to the left, the green curve is for $V=3$~TeV, the red one is for $V=2$~TeV, the blue is for $V=1.3$~TeV and the black curve is for $V=1$~TeV. The narrow shaded band encloses the allowed region for $0.096 < \Omega_{DM} h^2 < 0.122$ according to WMAP.}
\label{fig:4}
\end{figure}

\begin{figure}[h]
\centering
\includegraphics[width=0.99\columnwidth]{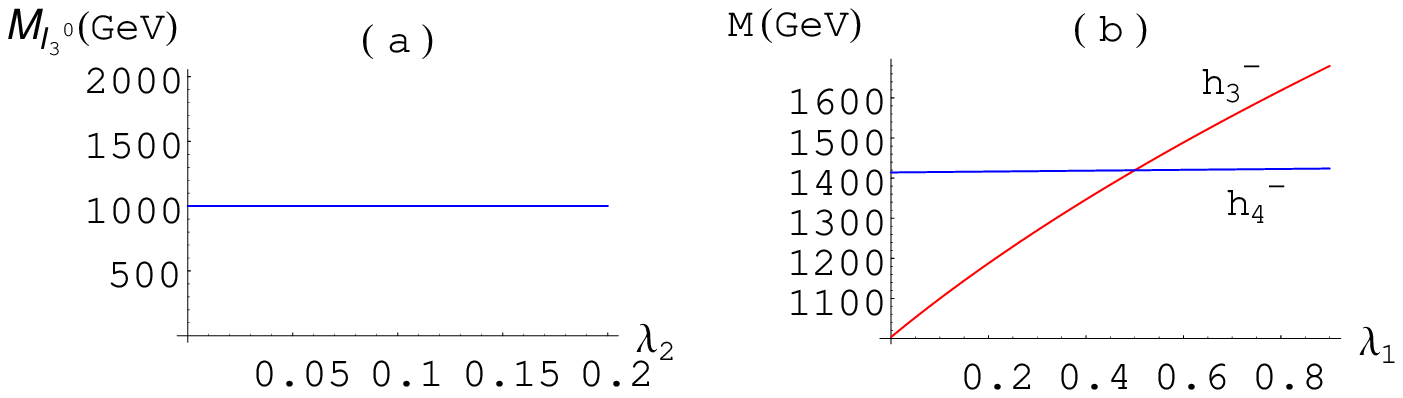}  \includegraphics[width=0.99\columnwidth]{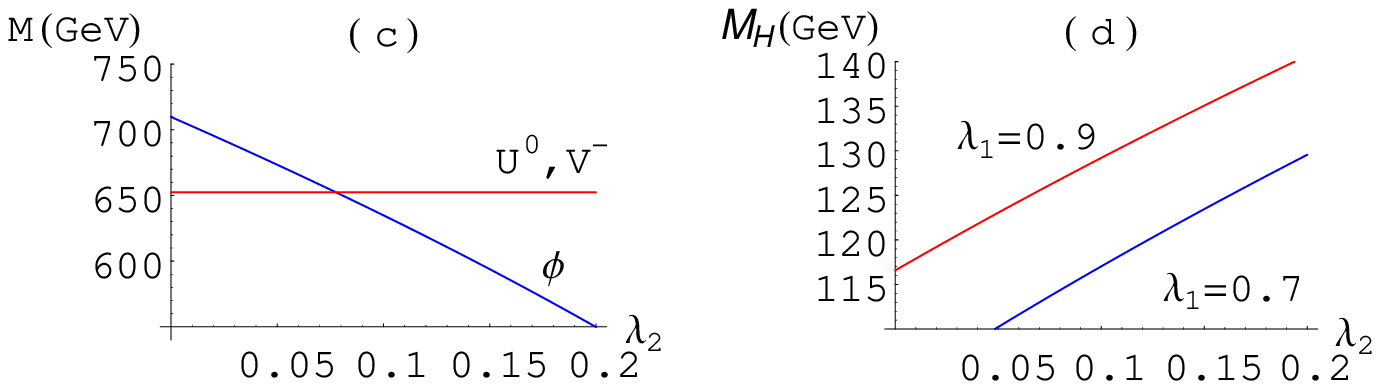}
\caption{Mass spectrum for the additional scalars and vector bosons of 3-3-1RH$\nu$ model for fixed $V=2$~TeV. In (a) we show the neutral heavy pseudo-scalar as a function of $\lambda_2$, for $\lambda_1=0.7$. In (b) we plot the charged scalar masses valid for all values of $\lambda_2$ as a function of $\lambda_1$. In (c) we show the WIMP mass against the vector bileptons masses as a function of $\lambda_2$, for $\lambda_1=0.7$. Finally, in (d) we plot the Higgs mass for two values of $\lambda_1=0.7$ and $0.9$ as a function of $\lambda_2$.}
\label{fig:5}
\end{figure}

Summing up all these contributions to the WIMP annihilation~\footnote{The expressions for the $a$ and $b$ parameters in Eq.~(\ref{thermalexpansion}) are too lengthy and we omit them in this work, though they can be easily obtained through the interactions listed in the appendix.}, we use equations (\ref{thermalexpansion}), (\ref{omega}) and (\ref{xf}) and impose the bound in Eq.~(\ref{limitsDM}) to see if there is any compatible range of parameters which gives the correct Dark Matter abundance. Our results are shown in Fig.~\ref{fig:4}.
In this figure we plotted the abundance against the coupling $\lambda_2$ for fixed $\lambda_1$, and the other couplings given in Eq.~(\ref{couplings}). For the range of values we took for the parameters, all scalars (except the Higgs) are always heavier than the WIMP, as shown in Fig.~\ref{fig:5}~\footnote{ Those plots were taken for $V=2$~TeV, but they possess a similar behavior for other values of $V$.}.
Since the Higgs mass is above $M_H=114$~GeV only for $\lambda_1 \geq 0.7$ (for $V = 2$~TeV), we used this value to get the abundance curves once the results are very weakly dependent on $\lambda_1$. From Fig.~\ref{fig:4} we observe that the larger the value of $V$, the larger is the maximum allowed value for $\lambda_2$. 

The curve for $V=1$~TeV presents only one valid region for $0.049\leq\lambda_2\leq 0.063$ leading to an acceptable dark matter abundance. However, for this value of $V$ the Higgs mass is below its experimental lower bound, $M_H > 114$~GeV, for $\lambda_1=0.7$. If we change this coupling to larger values, for example $\lambda_1=0.9$, the acceptable region is still discarded for $V=1$~TeV. This situation changes when we increase $V$ to a minimum value $V=1.3$~TeV, bringing the model to a comfortable position concerning the Higgs mass, even for $\lambda_1=0.7$. However, when we look to the WIMP mass, we conclude that also the $V=1.3$~TeV has problems for values of $\lambda_2 < 0.078$, since in this regime the vector bileptons would be lighter than our neutral scalar bilepton (see Fig.~\ref{fig:5}), jeopardizing its stability. This leaves only a tiny window for $\lambda_2$ in this case, $0.078\leq\lambda_2\leq 0.081$. For $V=2$~TeV and above, this problem is absent, though the possible range for $\lambda_2$ is still short. We stick with values for $V$ above $1.3$~TeV and not much bigger than $3$~TeV, in order to not reach the non-perturbative regime of 3-3-1 model, which would be attained at $V\approx 4$~TeV~\cite{alexlimit}.

It is appropriate to say that the above results were obtained for the bilepton quarks degenerated in mass, an assumption made to simplify our calculations, but departure of this should not modify our qualitative results. The only important point to be considered is that bilepton quarks should have a mass larger than $1.5$~TeV in order to generate sufficient abundance. This is a reasonable assumption since they receive mass from the largest scale in the model, $v_\chi^\prime = V$, which is in the TeV range. Larger values for the masses of bilepton quarks would just push the values of $\lambda_2$ to a bit larger values, without compromising our conclusions.

We should also remark that we limited our analysis to a very conservative scenario, where several parameters were fixed ad hoc, since they are almost free of constraints. This does not mean that the above results put severe constraints on the couplings of 3-3-1RH$\nu$ model if it has to contain a WIMP. We only meant to show that even with a restricted set of possibilities, the 3-3-1$RH\nu$ model offers a good WIMP candidate for CDM with no need of adjusting its parameters to extremely unnatural values. Besides, the 3-3-1$RH\nu$ WIMP has a very peculiar signature once it carries two units of lepton number and could be easily distinguished from other models like SUSY or Extra Dimensions in collider experiments.

From the above results, see Figs.~(\ref{fig:4}) and (\ref{fig:5}), our scalar bilepton 3-3-1RH$\nu$ model WIMP has a preferred mass around $M_\phi\approx 600$~GeV, which is about six times greater than that of neutralinos in the SUSY preferred scenario~\cite{DMmodels}, and about the same magnitude as the lightest neutral vector boson Kaluza-Klein first mode, $B^{(1)}$, in 5D Universal Extra Dimensions (UED)~\cite{Dobrescu}. 
The relative small WIMP mass in the case of SUSY can possibly be attributed to the fact that neutralinos are Majorana fermions 
and the main contribution to its annihilation occurs through P-wave into fermions. This leads to small cross sections requiring a smaller WIMP mass in order to annihilate more efficiently and give the correct CDM abundance. As for the $B^{(1)}$ in UED,  its cross section can be shown to be roughly temperature independent~\cite{Dobrescu}, which means it efficiently annihilates into light fermions through S-wave allowing larger WIMP masses. Also, if the lightest WIMP of UED is a spinless photon in 6D, annihilation into fermions is helicity suppressed and presents a preferred mass half the size of the $B^{(1)}$ case (see the last paper on Ref.~\cite{Dobrescu}).

Since $\phi$ in 3-3-1RH$\nu$ model is a scalar, it is not S-wave suppressed for annihilation into gauge and Higgs boson, but as in the case of spinless photon in 6D UED, its annihilation into light fermions is also helicity suppressed. Indeed, for low values of $V$, the main contribution to the thermally averaged cross section comes from the annihilation into gauge bosons and, for some range of parameters, also into a pair of top quarks. Actually, as we have seen above, $V$ has to be bigger than 1.3~TeV, given the parameter space we are considering in this work, for which top pair annihilation is as important as gauge bosons to produce efficient depletion of $\phi$ leading to the right CDM abundance. This happens because P-wave is enhanced since the top quark is heavy and we choose a rather strong $\phi-T-t$ coupling. But also because the bilepton quark $T$ exchanged in {\it t} and {\it u} channels is not extremely heavy, we took $M_T=1.5$~TeV, otherwise this contribution would be negligible and enough annihilation would occur for smaller $\phi$ mass. This is in agreement with our above results for $\Omega_{CDM}$, which would demand increasing values for $\lambda_2$ as $M_T$ increases.
Thus, it is possible that such a similarity between the size of WIMP mass in 3-3-1RH$\nu$ model and the vector boson in 5D UED is an artifact of the peculiarities of the chosen parameters in the former model.

\subsection{WIMP direct detection}
\label{subsec2}

After concluding that our proposed WIMP possesses some region of the parameter space which is in agreement with data on CDM abundance, we should at least check if some detection is possible for our candidate, or if there is any contradiction with the experimental exclusion limits put by the latest results on WIMP detection.
The most probable signal of a WIMP is expected to appear in direct detection experiments, which consists of measuring the recoil energy of nuclei when these are elastically scattered by a WIMP~\cite{DMmodels}. The data can be translated to a cross section normalized to nucleon and are usually presented as a limit on this cross section. We are going to analyze then the chance of direct detection  of our scalar bilepton WIMP through elastic scattering with nuclei using the CDMS and XENON collaboration data~\cite{wimpexperiments}, which are the most stringent current results on WIMP-nuclei elastic collision.  We will also use the projected sensitivities for WIMP detection in future experiments~\cite{SuperCDMS,xenon1t}.

\begin{figure}[h]
\centering
\includegraphics[width=0.9\columnwidth]{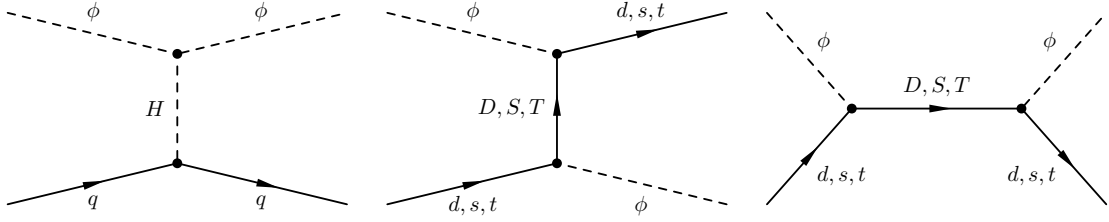}
\caption{Diagrams that contribute to the WIMP-nucleon elastic scattering cross section through WIMP-quark interactions. The first graph is the t channel Higgs exchange, while the remaining ones are s and t channel bilepton quarks exchange.}
\label{fig:6}
\end{figure}

Our WIMP interacts with nucleons through its couplings with quarks by exchanging a Higgs boson or a bilepton quark (see Fig.~\ref{fig:6}). 
The S matrix amplitude for these processes are given by,
\beq
{\cal M}_a &=& -i\sqrt{\frac{G_F}{\sqrt{2}}}\frac{m_q}{M_H^2}\left( 2\lambda_2 v_\eta +\lambda_6 v_\rho \right) {\bar q}q\,,\nonumber \\
{\cal M}_b &=& -i\frac{g_{qq^\prime}^2}{(M_\phi+E_q)^2-M_{q^\prime}^2}\left\{\frac{M_\phi+E_q}{2}\left[{\bar q}\gamma^0 q + {\bar q}\gamma^0 \gamma^5 q\right]\right\}\,,\nonumber \\
{\cal M}_c &=& -i\frac{g_{qq^\prime}^2}{(M_\phi-E_q)^2-M_{q^\prime}^2}\left\{\frac{M_\phi-E_q}{2}\left[{\bar q}\gamma^0 q + {\bar q}\gamma^0 \gamma^5 q\right]\right\}\,,
\label{amplitudes}
\eeq
where $G_F$ is the Fermi constant, $m_q$ is the SM quark mass, $M_{q^\prime}$ is the bilepton quark mass and the couplings in the last two amplitudes (bilepton quark exchange) read as,
$g_{qq^\prime}=g_{11}$ for $q=d$ and $q^\prime =D$, $g_{qq^\prime}=g_{22}$ for $q=s$ and $q^\prime =S$ and $g_{qq^\prime}=h_{33}$ for $q=t$ and $q^\prime =T$, no other quark contributes to these amplitudes. In the above equations we have used the non-relativistic limit, which also allows us to discard the term ${\bar u}(p_1)\gamma^0\gamma^5 u(p_2)\approx 0$. In this limit we can write $\langle {\bar q}\gamma^0 q\rangle\approx \langle {\bar q} q\rangle$. Besides, we can neglect the quark tri-momentum and write $E_q\approx m_q$.

As expected we will have only spin-independent (SI) contributions to the WIMP-nucleon cross section since our WIMP is a scalar. It is useful to define then the following average amplitude:
\be
\langle {\cal M}\rangle = i \alpha_q \langle {\bar q} q\rangle\,,
\label{ampli2}
\ee
whith $\alpha_q$ given by,
\beq
\alpha_q &=& -\left\{\left(\frac{G_F}{\sqrt{2}}\right)^\frac{1}{2}\frac{m_q}{M_H^2}(2\lambda_2 v_\eta + \lambda_6 v_\rho )+  \frac{g_{qq^\prime}^2}{(M_\phi + m_q)^2-M_{q^\prime}^2}\frac{M_\phi+m_q}{2} \right. \nonumber \\
&& \left. +  \frac{g_{qq^\prime}^2}{(M_\phi - m_q)^2-M_{q^\prime}^2}\frac{M_\phi-m_q}{2}  \right\}\,.
\label{alphaq}
\eeq

We write the matrix elements for the quarks in the nucleon by separating the contributions of light and heavy quarks as follows~\cite{DMmodels},
\beq
\langle {\bar q} q\rangle &=& \frac{m_{p,n}}{m_q}f_{T_q}^{p,n}\,\,\,\,\,\,\,\,\,\,\,\,\,\,\mbox{ for  u, d, s}\,, \nonumber \\
\langle {\bar q} q\rangle &=& \frac{2}{27}\frac{m_{p,n}}{m_q}f_{T_G}^{p,n}\,\,\,\,\,\,\,\mbox{ for  c, b, t}\,.
\label{quarksnucl}
\eeq
Here the subscripts $p$ and $n$ label the proton and the neutron, respectively.

In order to obtain the WIMP-nucleon coupling it remains to sum over the quarks,
noticing that all the six SM quark flavors contribute in Higgs exchange channel, while in the bilepton quarks exchange channels only down, strange and top quarks contribute, as can be seen from Eq.~(\ref{alphaq}). We then get,
\be
f_{p,n}^\phi = m_{p,n}\sum_{q=u,d,s}\frac{\alpha_q}{m_q}f_{T_q}^{p,n} +\frac{2}{27}m_{p,n}f_{T_G}^{p,n}\sum_{q=c,b,t}\frac{\alpha_q}{m_q}\,,
\label{wimpnucl}
\ee
where $f_{T_u}^p=0.020\pm 0.004$, $f_{T_d}^p=0.026\pm 0.005$, $f_{T_s}^p=0.118\pm 0.062$, $f_{T_u}^n=0.014\pm 0.003$, $f_{T_d}^n=0.036\pm 0.008$, $f_{T_s}^n=0.118\pm 0.062$ (see Ref.~\cite{fs}) and $f_{T_G}^{p,n}$, which is due to the $\phi$ coupling to gluons through loops of heavy quarks, is obtained from, 
\be
f_{T_G}^{p,n}=1-\sum_{q=u,d,s}f_{T_q}^{p,n}\,,
\ee
leading to the values, $f_{T_G}^{p}\approx 0.84$ and $f_{T_G}^{n}\approx 0.83$.

The WIMP-nucleus SI elastic cross section results from summing the nucleons in the target, yielding, at zero momentum transfer,
\be
\sigma_0 = \frac{m_N^2}{4\pi (M_\phi + m_N)^2}\left( Z f_p^\phi + (A-Z) f_n^\phi \right)^2\,,
\label{wimpnucleo}
\ee
with, $m_N$ the nucleus mass, $Z$ the atomic number and $A$ the atomic mass.
What is usually employed in expressing constraints on WIMP-nuclei elastic experiments is the WIMP-nucleon cross section, which in the SI case reads,
\be
\sigma_{p,n}^{SI} = \sigma_0 \frac{m_{p,n}^2}{m_r^2 A^2}\,,
\label{wimpnucleoncs}
\ee
where $m_r=M_\phi m_N/(M_\phi + m_N)$ is the reduced WIMP mass.

\begin{figure}[h]
\centering
\includegraphics[width=0.7\columnwidth]{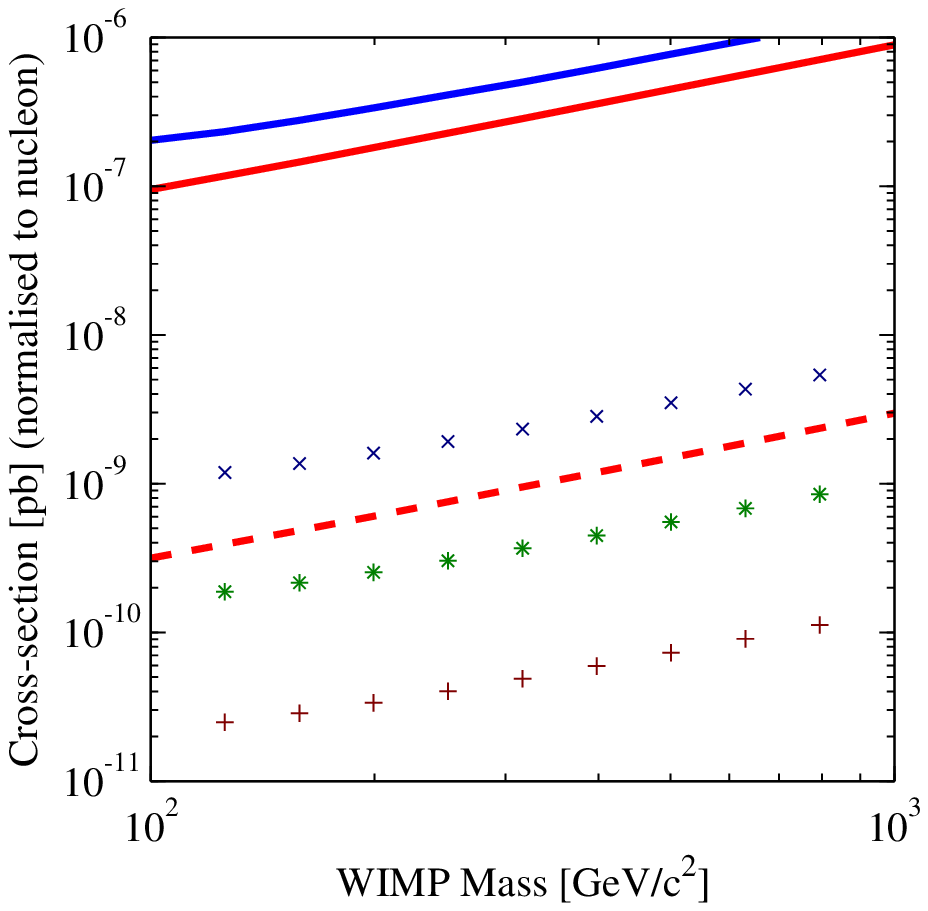}
\includegraphics[width=0.7\columnwidth]{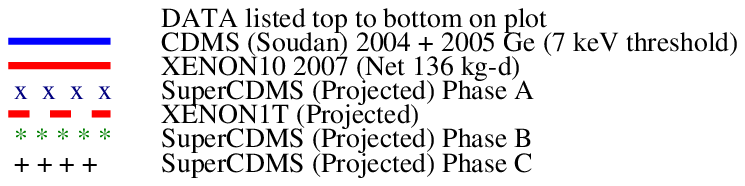}
\caption{Spin-independent WIMP-nucleon elastic scattering cross section sensitivity (current and projected) from CDMS and XENON Collaborations.}
\label{fig:7}
\end{figure}

\begin{figure}[h]
\centering
\includegraphics[width=0.99\columnwidth]{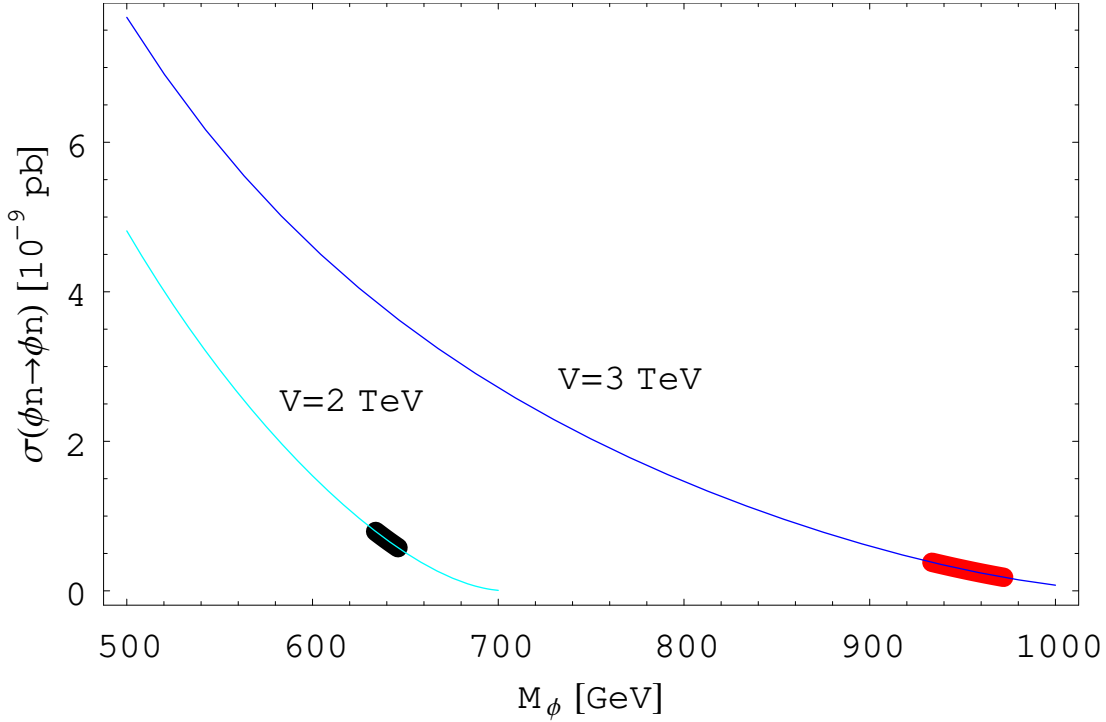}
\caption{Spin-independent WIMP-nucleon cross sectionof the 3-3-1$RH\nu$ model as a function of the WIMP mass.The upper curve is for $V=3$~TeV and the lower one is for $V=2$~TeV, both taken for $M_{q^\prime}=1.5$~TeV. The thicker regions on these lines are those in agreement with the bounds on $\Omega_{CDM}$ imposed by WMAP.}
\label{fig:8}
\end{figure}

Finally, we can compute this normalized cross section for the $^{73}$Ge (appropriate for CDMS spin-independent cross section) using the values of the 3-3-1RH$\nu$ model parameters fixed as before.
In Fig.~\ref{fig:7} we present the current and projected data for WIMP-nucleon cross section from CDMS and XENON Collaborations~\cite{wimpexperiments,SuperCDMS,xenon1t,berkeley}, ans in Fig.~\ref{fig:8} we present our results from 3-3-1RH$\nu$ model.
In Fig.~\ref{fig:8} we have plotted only the results for $V=2$~TeV and $V=3$~TeV for convenience, where
the thick regions are the allowed ones for WMAP bounds on $\Omega_{CDM}$. However, the reader should have in mind that what we really have is a continuous range of WIMP mass and cross-section, which would fill a band between those two small regions shown in the figure.
As we can see from Figs.~\ref{fig:7} and \ref{fig:8},  $\phi$ is close to the threshold detection only for projected sensitivity of XENON1T experiment~\cite{xenon1t}, but still below this threshold for the whole range of parameters assessed here. 
However, for this range of parameters, $\phi$ direct detection might be realized at improved sensitivity, lying between the projected phases B and C of Super-CDMS~\cite{SuperCDMS}, covering WIMP-nucleon cross-sections a little bigger than $10^{-11}$~pb. 

In the case of neutralinos in SUSY and $B^{(1)}$ in 5D UED, there is also a spin-dependent contribution to be considered, which is dominant when $A\leq 20$, but suppressed for bigger values of $A$. For 6D UED spinless photon, similar to our case, there is only SI WIMP-nucleon scattering. The SI cross sections should differ basically for specific model dependent WIMP-quark couplings. Concerning neutralinos, typical values for this SI normalized cross section is $\sigma_{p,n}^{SI}\approx 10^{-12}-10^{-6}$~pb, while for $B^{(1)}$ in 5D UED $\sigma_{p,n}^{SI}\approx 10^{-10}$~pb for WIMP mass about 1~TeV. In the case of spinless photon in 6D UED, $\sigma_{p,n}^{SI}\approx 10^{-11}-10^{-9}$~pb, considering the region of WIMP mass for which the observed amount of CDM is generated, $M \approx 200 - 300$~GeV. In our model, only $d$, $s$ and $t$ quarks participate in WIMP-nucleon interaction, but this does not affect appreciably the amount of WIMP-nucleon scattering, yielding similar results as the UED model, namely $\sigma_{p,n}^{SI}\approx 10^{-11}-10^{-10}$~pb for $M_\phi\approx 600 - 1000$~GeV.  

Also, collider signatures may be pursued for the forthcoming LHC at CERN~\cite{LHC}, which will have enough center of mass energy to test several new particles at TeV scale. Collider signatures and indirect detection~\cite{indirect}, which concerns WIMP annihilation  into SM particles like photons and neutrinos, should be carried out in the future when more constraints on the parameter space of 3-3-1RH$\nu$ model are available. Nevertheless, our aim here was to show that a scalar bilepton WIMP can be realized in 3-3-1RH$\nu$ model, providing a completely distinct candidate for explaining CDM, reproducing its observed relic abundance and in agreement with the most stringent constraints from direct detection experiments. It is true that we have investigated just a tiny range of the parameter space, but this limitation does not invalidate our conclusions and can be better explored as long as we improve our knowledge concerning the phenomenological consequences of this model.

\section{Conclusions}
\label{sec5}

The 3-3-1RH$\nu$ model admits a couple of bilepton particles in its spectrum, raising the possibility of having a  CDM candidate, since bileptons carry two units of lepton number. This is so because such a very specific quantum number is appropriate to forbid its interaction with many of the electroweak fields, since they are allowed to decay only on other bileptons, which have to be heavier than SM particles.
Considering this scenario we obtained the particle mass spectrum of scalars in 3-3-1RH$\nu$ model and, by assuming some conditions over the parameter space, we have shown that the lightest bilepton in the model turns out to be a scalar, a combination of two scalar interaction eigenstates that we called $\phi$. 
For the region where the values of the parameters guarantee the stability of this scalar, we computed the $\phi$ abundance and obtained stringent constraints for the parameters in order to have agreement with WMAP results for CDM abundance. 
We have found that $\phi$ can have mass ranging from about $600$~GeV to some Few TeV, characterizing it as a heavy WIMP.
It is opportune to say that, although we have restricted our parameter space due  to lack of knowledge on several couplings in the model, we had no need to unnaturally adjust them to very small values as generally happens in several models, including Supersymmetry. In fact we assumed that these couplings are close to one, and checked that $\phi$ is an excellent candidate to represent a WIMP and explain the presently observed CDM abundance.

We also studied the possibility of observing this WIMP in direct detection experiments. For this we  have computed the elastic scattering $\phi$-nucleon cross section and contrasted our results with present and future experiments. We have seen that $\phi$ is still far from the range of detection for current and near future CDMS and XENON sensitivities, at least for the short parameter space considered in this work. However, even this limited scenario can be at reach for projected  Phases B and C of Super-CDMS~\cite{SuperCDMS}.
Besides this, it would be interesting to pursue the production of $\phi$ at collider experiments, mainly at LHC, and also extend our search including a larger region of the parameter space considering additional phenomenological constraints on 3-3-1RH$\nu$ model from Collider physics and Cosmology, as well as include prospects for $\phi$ indirect detection too, a gap we wish to fill soon. 

Finally, we would like to stress that our proposed WIMP is not only feasible but a reasonable alternative in the sense that the Particle Physics model we are dealing with is only a small extension of the SM gauge group, whose scale is about to be assessed at LHC. There are several features that distinguishes the 3-3-1RH$\nu$ model from other extensions, like SUSY and Extra dimensions models. Namely, we have not only bilepton scalars in the spectrum but vector bosons and quark bileptons, all of them acquiring mass at hundreds of GeV. Certainly their signal at detectors are worth to be studied. Besides, new phenomena are predicted in this model~\cite{331valle,lightnu,axionmajoron,SCPV}, including neutrinoless double beta decay, rare decays, new sources of CP violation and so on. Presence of such signals  would reinforce our expectation concerning a bilepton WIMP to explain CDM in the Universe.
\\ \\

\noindent {\bf Acknowledgments:}\\
The authors acknowledge the support of the Conselho Nacional de Desenvolvimento Cient\'{\i}fico e Tecnol\'ogico (CNPq).

\section*{Appendix A}
\renewcommand{\theequation}{A.\arabic{equation}}
\setcounter{equation}{0}
In this Appendix we show some interactions of  interest for the computation of the WIMP abundance.
\begin{itemize}
{\item Interactions between WIMP and gauge bosons
\begin{eqnarray}
	\frac{g^2 v_\eta}{2\sqrt{2}} \phi W^+_\mu V^{- \mu}+ \mbox{H.c}\,;
\end{eqnarray}
}
\end{itemize}
\begin{itemize}
{\item Interactions between WIMP and the  Higgs
\begin{eqnarray}
	-\frac{1}{\sqrt{2}}(2\lambda_2v_\eta +\lambda_6 v_\rho)H \phi^{*}\phi - 
	\frac{1}{\sqrt{2}}(\lambda_2 +\frac{\lambda_6}{2})H H \phi^{*}\phi +
	\mbox{H.c}\,;
\end{eqnarray}
}
\end{itemize}
\begin{itemize}
{\item Interactions between WIMP and the  quarks
\begin{eqnarray}
	-g_{ia}\bar{d^\prime}_{iL}d_{aR}\phi^* - h_{3a}\bar{u^\prime}_{3L}u_{aR}\phi +
	\mbox{H.c}\,;
\end{eqnarray}
}
\end{itemize}
\begin{itemize}
{\item Interactions between Higgs and the gauge bosons
\begin{eqnarray}
	-\sqrt{2}g^2(v_\eta +v_\rho)H W^+_\mu V^{- \mu} + \mbox{H.c}\,;
\end{eqnarray}
}
\end{itemize}
\begin{itemize}
{\item Interactions between Higgs bosons
\begin{eqnarray}
	\frac{1}{2\sqrt{2}}g^2((\lambda_2+\lambda_6)v_\eta +\frac{\lambda_6}{2}v_\rho)H H H + \mbox{H.c}\,;
\end{eqnarray}
}
\end{itemize} 
\vskip 1cm
%
\noindent{\bf References} \\

\end{document}